\documentclass[aps,nofootinbib,preprint]{revtex4}

\usepackage{graphicx}

\usepackage{amsmath}
\usepackage{amsfonts}
\usepackage{amssymb}
\usepackage{hyperref}
\usepackage{color}

\newcommand{\be}{\begin{equation}}
\newcommand{\ee}{\end{equation}}

\def\beq{\begin{equation}}
\def\eeq{\end{equation}}
\def\bea{\begin{eqnarray}}
\def\eea{\end{eqnarray}}
\def\ov{\overline}

\def\slashchar#1{\setbox0=\hbox{$#1$}           
   \dimen0=\wd0                                 
   \setbox1=\hbox{/} \dimen1=\wd1               
   \ifdim\dimen0>\dimen1                        
      \rlap{\hbox to \dimen0{\hfil/\hfil}}      
      #1                                        
   \else                                        
      \rlap{\hbox to \dimen1{\hfil$#1$\hfil}}   
      /                                         
   \fi}                                        %

\begin{document}


\vspace*{2cm}
\title{Goldstini as the decaying dark matter}

\author{\vspace{0.5cm}Hsin-Chia Cheng$\,^a$, Wei-Chih Huang$\,^{b,c}$, Ian Low$\,^{b,c}$, and Arjun Menon$\,^d$}
\affiliation{\vspace{0.5cm}
\mbox{$^a$Department of Physics, University of California, Davis, CA 95616}\\
 \mbox{$^{b}$High Energy Physics Division, Argonne National Laboratory,
Argonne, IL 60439}\\
\mbox{$^{c}$Department of Physics and Astronomy, Northwestern University,
Evanston, IL 60208}\\
\mbox{$^{d}$Illinois Institute of Technology, Chicago, IL 60616}
\vspace{0.8cm}
}

\begin{abstract}
\vspace*{0.5cm}

We consider a new scenario for supersymmetric decaying dark matter without $R$-parity violation in  theories with goldstini, which arise if supersymmetry is broken independently by multiple sequestered sectors. The uneaten goldstino naturally has a long lifetime and decays into three-body final states including the gravitino, which escapes detection, and two visible particles. The goldstini low-energy effective interactions are derived, which can be non-universal and allow the dark matter to be leptophilic, in contrast to the case of a single sector supersymmetry breaking. In addition, the three-body decay with a missing particle gives a softer spectrum. Consequently, it is possible to fit both the $e^+/e^-$ excess observed by the PAMELA and the $e^++e^-$  measurements by the Fermi-LAT using universal couplings to all three lepton flavors or 100\% branching fraction into electrons/positrons, both of which are disfavored in the conventional scenario of dark matter decays into two or four visible particles without missing energy.

\end{abstract}


\maketitle

\section{Introduction}
\label{sect:intro}

The existence of the dark matter has been firmly established and it constitutes about 23\% of the total energy density in the universe. The nature of the dark matter is one of the most outstanding questions in cosmology and particle physics. Many different types of experiments are deployed to detect the dark matter and to measure its properties, including direct detections from the recoils of the nuclei hit by the dark matter particle, indirect detections of the  cosmic rays from dark matter annihilations or decays, and collider searches by direct production of dark matter particles. Recently, there has been an interesting observation of anomalous $e^+/e^-$ excess in the energy range of 1 -- 100 GeV measured by the PAMELA collaboration \cite{Adriani:2008zr}, which may be interpreted as indirect dark matter signals, coming from dark matter annihilations or decays inside the galactic halo. In addition, the $e^+ +e^-$ spectrum measured by the Fermi-LAT experiment between 20 GeV and 1 TeV is harder than that inferred from previous experiments~\cite{Abdo:2009zk}, which may also be attributed to the contribution from the dark matter. In this paper, we consider a new dark matter candidate which could naturally produce the excess of the electron/positron flux observed in these experiments.

The cosmic positrons are one of the prominent signals for indirect dark matter detections. For the most popular dark matter candidate, a weakly interacting massive particle (WIMP), electrons and positrons can be produced from annihilations of the WIMPs in the galactic halo. However, to account for the PAMELA excess, a large boost factor at the order of 100 or larger is required to increase the annihilation rate~\cite{Bergstrom:2008gr,Barger:2008su,Cholis:2008hb,Cirelli:2008pk}. In addition, large flux of gamma rays will be produced in the dark matter annihilations, which is severely constrained by the observed gamma ray spectrum~\cite{Bertone:2008xr,Nardi:2008ix,Essig:2009jx,Meade:2009iu}. As a result, explaining the PAMELA excess by annihilating dark matter has a hard time to satisfy the constraints from the annihilation cross section and cosmic gamma ray data. Another possibility to generate the observed electron/positron spectrum is that if the dark matter particle is not absolutely stable, but decays with a very long lifetime. A small fraction of the dark matter particles has decayed, producing electrons and positrons in the decay products~\cite{Chen:2008dh}. The decaying dark matter has an easier time to satisfy the gamma ray constraints, but to explain the PAMELA excess the lifetime needs to be of the order of $10^{26}$--$10^{27}$ seconds~\cite{Nardi:2008ix,Essig:2009jx,Meade:2009iu}, which seems to be an additional arbitrary parameter coming from nowhere.

For the decaying dark matter, it is usually assumed that the symmetry that protects the stability of the dark matter particle is not exact, but violated by some highly suppressed interactions. It has been argued that the required lifetime can be obtained from a TeV scale particle decaying through dimension-6 operators suppressed by the grand unification scale $m_{\rm GUT} \sim 2\times 10^{16}$ GeV \cite{Nardi:2008ix, Arvanitaki:2008hq}. In this paper we consider another possibility that an exact symmetry is carried by two sequestered sectors, which interact indirectly only through the visible sector (standard model). The lightest particle charged under this symmetry is absolutely stable. However, the dark matter is made of the next to the lightest particle charged under the same symmetry, which is only approximately stable due to sequestering. The dark matter particle decays to the truly stable particle with a long lifetime because of the highly suppressed interactions between the two sequestered sectors. The standard model (SM) particles produced in the decays can be observed, and could be responsible for the anomalies in the cosmic ray experiments. 

We show that such decaying dark matter can arise naturally in the goldstini scenario proposed recently~\cite{Cheung:2010mc}. In this scenario, supersymmetry (SUSY) is spontaneously broken in more than one sequestered sectors. There is a goldstino associated with the spontaneously broken SUSY in each sector. The SUSY in different sectors are connected by supergravity and only one combination of the goldstini is eaten and becomes the longitudinal mode of the gravitino. The other combinations of the goldstini acquire a mass of twice the gravitino mass at the lowest order due to the supergravity effect. Assuming $R$-parity is exactly conserved, and if the gravitino and an uneaten goldstino are the lightest and the next to the lightest supersymmetric particles (LSP and NLSP), respectively, the cosmic electrons and positrons can be produced from decays of the goldstino dark matter to the gravitino. If the two SUSY breaking sectors only interact indirectly through the visible supersymmetric standard model (SSM) sector, the interactions responsible for the goldstino decays are highly suppressed and the required lifetime for the observed electron/positron excess can be naturally obtained. 

A distinct feature of this scenario is that the dark matter decays dominantly through three-body processes, producing a pair of SM particles and another invisible massive particle. Most of the studies of decaying dark matter before assumed that the dark matter particle decays through a two-body process to a pair of SM particles or a pair of portals to four SM particles without additional missing particles. Some exceptions are in Ref.~\cite{Ibarra:2009dr} where the three-body decays including a neutrino, as well as from internal bremsstrahlung, are considered. The constraint on the anti-proton flux, which shows no excess in the PAMELA experiment \cite{Adriani:2008zq}, requires that the decays of dark matter particles dominantly produce leptons. In the case of two-body decays, the muon and tau final states are preferred \cite{Meade:2009iu} and the direct decay to the electron and positron pair would give a sharp edge on the energy spectrum at half the mass of the dark matter particle, which is not seen by Fermi-LAT. On the other hand, the electrons and positrons coming from the three-body decays will have a softer and smooth spectrum which may still be consistent with other observations. As will be shown, the goldstino couplings to the SM particles, unlike the universal coupling of the gravitino, are governed by the fractions of the soft-SUSY breaking masses coming from different SUSY breaking sectors for the corresponding superpartners. It is easy for the goldstino to have preferential decays to leptons if different superpartners receive different soft masses from different SUSY-breaking sectors.

This paper is organized as follows. In Sec.~\ref{sect:effgold} we derived the goldstini interactions with SM fermions using the method of constrained superfields developed recently by Komargodski and Seiberg~\cite{Komargodski:2009rz}.  From the interactions we can calculate the decay rate of a goldstino to the gravitino and a pair of SM fermions. The interactions with other SM fields are collected in the Appendix. In Sec.~\ref{sect:decayingDM} we discuss the model of the decaying goldstino dark matter and the parameters which can give rise to the PAMELA signal and satisfy other astrophysical and cosmological constraints. In Sec.~\ref{sect:pheno} we perform fits to the electron/positron energy spectra observed by PAMELA and Fermi-LAT experiments with the decaying goldstino scenario, and identify decay modes and parameters which can be consistent with the observation data. We then briefly discuss the collider phenomenology of this scenario. Conclusions are drawn in Sec.~\ref{sect:conclusions}. Throughout this paper, we use ``goldstini'' when we refer to the goldstino fields coming from different SUSY-breaking sectors, and ``goldstinos'' to represent the plural form of the same-species goldstino.

\section{Effective interactions of goldstini}
\label{sect:effgold}

In this section we derive the low-energy effective interactions of goldstini with two standard model fermions, by  the method of constrained superfields introduced in Ref.~\cite{Komargodski:2009rz}. In order to highlight the non-universal nature of goldstini interactions,  we begin by reviewing the low-energy effective interactions of one goldstino.

Ref.~\cite{Komargodski:2009rz} considers a system of two chiral superfields $X$ and $Q$,
\be
X = \tilde{x} + \sqrt{2} \theta \eta + \theta^2 F_X \ , \qquad Q = \tilde{q} + \sqrt{2} \theta q + \theta^2 F_Q \ ,
\ee
interacting through the following K\"ahler potential $K$ and superpotential $W$
\be
K= X \ov{X} + Q\ov{Q} -\frac{c}{\Lambda^2} X^2 \ov{X}^2 - \frac{\hat{c}}{\Lambda^2} Q\ov{Q} X\ov{X} \ , \qquad
W = f X \ .
\ee
SUSY is spontaneously broken by the $F$-term vacuum expectation value (VEV) of $X$, where the goldstino resides, while $Q$ is the generic matter field such as the quark or the lepton. The $c$ and $\hat{c}$ terms are included in the K\"ahler potential to lift the unwanted massless scalars. The lagrangian right below the scale $\Lambda$ is given by
\be
\label{eq:uvL}
{\cal L}= \int d^4 \theta\, K + \int d^2\theta \, W + \int d^2\ov{\theta} \, \ov{W}  \ .
\ee
We are interested in finding the interactions at energies much below the soft SUSY-breaking mass scale $m_{soft}$,
\be
E \ll m_{soft} \ll \Lambda \ ,
\ee
where the scalar components of $X$ and $Q$ are integrated out. The zero-momentum lagrangian is given by
\be
\label{eq:1scalar}
{\cal L} =  -f^2 + |F_X + f|^2+ |F_Q|^2 -\frac{c}{\Lambda^2} \left|2\tilde{x} F_X -\eta^2 \right|^2 - \frac{\hat{c}}{\Lambda^2} \left|\tilde{q} F_X + \tilde{x} F_Q - q \eta \right|^2 \ ,
\ee
which gives rise to the following equations of the motion:
\bea
&&  \tilde{q} F_X + \tilde{x} F_Q - \eta\, q = 0 \ , \\
&&  2 \, \tilde{x} F_X -\eta^2 = 0 \ ,
\eea
The solutions turn out to be independent of the non-renormalizable couplings $c$ and $\hat{c}$.  After substituting the solutions back into the the chiral superfields, we obtain the {\em constrained} superfields:
\bea
X_{NL} &=&  \frac{\eta^2}{2F_X} + \sqrt{2} \theta \eta + \theta^2 F_X \ , \\
Q_{NL} &=& \frac{q \eta}{F_X} - \frac{\eta^2}{2F_X^2}F_Q + \sqrt{2}\theta q + \theta^2 F_Q \ .
\eea
These two superfields satisfy the constraints:
\be
X_{NL}^2 = 0 \ , \qquad Q_{NL}\, X_{NL} = 0 \ .
\ee
Because the zero-momentum lagrangian in Eq.~(\ref{eq:1scalar}) vanishes when evaluated at the solutions to the equations of motion, the leading effective interactions involving two goldstinos and two $q$'s  are obtained from the kinetic term of $\tilde{q}$:\footnote{There is another operator of the form, $(q \sigma_\nu\bar{q})(\partial_\mu \eta \sigma^\nu \partial^\mu\ov{\eta})$, which is subleading and only generated at the loop-level \cite{Komargodski:2009rz}.}
\be
{\cal L}_{eff} = \frac1{f^2}\partial_\mu( \ov{\eta}\, \ov{q}) \partial^\mu (\eta \, q)  + \cdots \ .
\ee
Evidently, the interaction is universal in flavors and only depends on the SUSY-breaking scale $f$.

Next we consider the ``goldstini'' scenario \cite{Cheung:2010mc} where SUSY is broken independently by two sequestered sectors. Matter fields in the SSM may interact with the two SUSY breaking sectors only via higher-dimensional operators suppressed by $\Lambda_1$ and $\Lambda_2$, respectively. The K\"ahler potential and superpotential in this case are
\be
\label{eq:goldkahler}
K = \sum_{i=1,2}\left( X_i \ov{X}_i    -\frac{c_i}{\Lambda^2} X_i^2 \ov{X}_i^2 - \frac{1}{\Lambda_i^2} X_i\ov{X}_i Q\ov{Q}\right)  + Q \ov{Q} \ , \qquad
W = \sum_{i=1,2} f_i X_i \ .
\ee
The form of the superpotential determines the combination eaten by the gravitino $\tilde{G}$ via the super-Higgs mechanism. The eaten goldstino and the uneaten orthogonal combination are related to the goldstini of the two sectors by
\be
\begin{pmatrix} \eta_1 \\
                            \eta_2 \end{pmatrix} =
\begin{pmatrix} \cos\theta & -\sin\theta  \\
                            \sin\theta & \cos\theta  \end{pmatrix}
\begin{pmatrix} \widetilde{G}_L \\
                            \zeta \end{pmatrix}    \ ,
\label{goldstinimassbasis}
\ee
where $\widetilde{G}_L$ is the longitudinal component of the gravitino, and we define
\be
\tan \theta = \frac{f_2}{f_1} \ , \qquad f_{eff} = \sqrt{f_1^2 + f_2^2} \ .
\ee
To derive the effective interactions at energies much below $m_{soft}$, we follow the same procedure as in the single goldstino case to integrate out the scalar components in $X_i$ and $Q$. Furthermore, since we are only interested in the leading-order contribution, the computation is greatly simplified if we replace all $F$-terms by their respective VEV's. In the end the zero momentum lagrangian is
\be
\label{eq:Lscalar}
{\cal L} = \sum_{i=1,2} \left( -f_i^2 + |F_i + f_i|^2 -\frac{c_i}{\Lambda^2} \left|2\tilde{x}_i F_i -\eta_i^2 \right|^2 - \frac1{\Lambda_i} \left|\tilde{q} F_i + \tilde{x}_i F_Q - q \eta_i \right|^2 \right)\ , 
\ee
and solutions to the equations of the motions for the scalars are
\bea
\tilde{x}_1 &=& \frac{\eta_1^2}{2f_1^2} \ , \qquad
 \tilde{x}_2 = \frac{\eta_2^2}{2f_2^2} \ , \\
 \tilde{q} &=& \frac{1}{{f_1^2}/{\Lambda_1^2}+{f_2^2}/{\Lambda_2^2}}\left(\frac{f_1}{\Lambda_1^2}\, \eta_1 q + \frac{f_2}{\Lambda_2^2}\, \eta_2 q\right) \nonumber \\
&=& \frac1{f_{eff}} \left[ \widetilde{G}_L\, - \left(\frac{\widetilde{m}_1^2 \tan \theta - \widetilde{m}_2^2 \cot \theta}{m_{\widetilde{q}}^2}\right) \zeta
\right] \,q ,
 \eea
where $\widetilde{m}_i^2 = f_i^2/\Lambda_i^2$ is the contribution from each SUSY-breaking sector to the scalar mass of $Q$ and $m_{\widetilde{q}}^2 \equiv \widetilde{m}_1^2 + \widetilde{m}_2^2$.

It turns out that there are two contributions to four-fermi interactions, in contrast to the case of a single goldstino. The first one comes from substituting the solution back into the lagrangian in Eq.~(\ref{eq:Lscalar}):
\be
 {\cal L}_{2f}^{(0)}= \frac{f_{eff}^2}{f_1^2\Lambda_2^2+f_2^2\Lambda_1^2} \,\ov{\zeta} \ov{q}\, \zeta q = \frac1{m_{\widetilde{q}}^2}\left(\frac{\widetilde{m}_1^2}{\Lambda_2^2} + \frac{\widetilde{m}_2^2}{\Lambda_1^2}\right)  \,\ov{\zeta} \ov{q}\, \zeta q \ ,
\ee
while the second one originates from the scalar kinetic term, $\partial_\mu \widetilde{q}^\dagger \partial^\mu \tilde{q}$, which is derivatively coupled:
\bea
\label{eq:Leff}
{\cal L}_{2f}^{(1)} &=& \frac1{f_{eff}^2} \partial_\mu (\ov{\widetilde{G}_L} \, \ov{q}) \partial^\mu (\widetilde{G}_L \, q) + \frac1{f_{eff}^2}\left(\frac{\widetilde{m}_1^2 \tan \theta - \widetilde{m}_2^2 \cot \theta}{m_{\widetilde{q}}^2}\right)^2
\partial_\mu (\ov{\zeta}\ov{q}) \partial^\mu (\zeta q) \nonumber \\
&& -\frac1{f_{eff}^2}\left(\frac{\widetilde{m}_1^2 \tan \theta - \widetilde{m}_2^2 \cot \theta}{m_{\widetilde{q}}^2}\right)
 \partial_\mu (\ov{\zeta}\ov{q}) \partial^\mu (\widetilde{G}_L q) + {\rm h.\, c.} \ .
\eea
Notice that while ${\cal L}_{2f}^{(0)}$ is not derivatively coupled, it only involves the uneaten goldstino and not the gravitino. In this sense $\zeta$ is really a pseudo-goldstino. If we are only interested in the decay of the goldstino, ${\cal L}_{2f}^{(0)}$ obviously does not contribute. For the goldstino annihilations or scatterings, however, both contributions are equally important since the two derivatives in ${\cal L}_{2f}^{(1)}$ pull out two factors of order $m_{soft}$ and we have $m_{soft}^2/f_{eff}^2 \sim 1/\Lambda_i^2$.

Concentrating on ${\cal L}_{2f}^{(1)}$, which is relevant for decays of the goldstino, we see that the interaction involving only the gravitino is still flavor-universal and insensitive to higher dimensional operators in the K\"ahler potential in Eq.~(\ref{eq:goldkahler}), while those involving the uneaten goldstino are non-universal and do depend on details of the ultraviolet physics. Using this effective lagrangian we can compute the decay width of the goldstino into two standard model fermions plus the gravitino:
\bea
\label{eq:zgff}
\Gamma_{\zeta \to \widetilde{G}_L f\bar{f}} = \frac{N_c m_\zeta^9}{15360\pi ^3 f_{\rm eff}^4}
\left(\frac{\widetilde{m}_1^2 \tan \theta - \widetilde{m}_2^2 \cot \theta}{
m_{\widetilde{q}}^2}\right)^2 F_f(x) \ ,
\eea
where $N_c=3$ for quarks and 1 for leptons, $x = m_{\widetilde{G}_L}/m_\zeta$, and
\bea
F_f(x) &=& (1-x^2) \left(2 x^{10}+x^9-6 x^8+6 x^7+4 x^6+106 x^5+4 x^4+6 x^3-6 
   x^2+x+2\right) \nonumber \\
& & +60 \left(x^7+x^5\right) \log x^2\ .
\eea
For the benchmark scenario in Ref.~\cite{Cheung:2010mc},  $m_\zeta=2m_{\widetilde{G}_L}$ and $F_f(1/2)\approx 0.8$.

The effective interaction with two fermions is the most relevant one for the purpose of this study. However, for completeness, we present effective interactions of the goldstini with other SM particles, such as the gauge bosons and the Higgs fields, in the Appendix.

\section{The Decaying Goldstino Dark matter}

\label{sect:decayingDM}

{}From the effective interactions derived in Section \ref{sect:effgold} we see that if the gravitino is the LSP and the goldstino the NLSP, the goldstino only decays through dimensions-8 operators. It can be cosmologically stable and play the role of the dark matter~\cite{Cheung:2010mc}. However, since the goldstino is not absolutely stable, a small fraction of its relic could have decayed and gives rise to interesting astrophysical signals. As mentioned in the Introduction, the recent anomalous $e^+/e^-$ excess measured by the PAMELA experiment may be interpreted as indirect dark matter signals, coming from dark matter annihilations or decays inside the galactic halo. The decaying dark matter has an easier time to satisfy the gamma ray constraints but requires the lifetime to be of the order of $10^{26}$--$10^{27}$ seconds. This long lifetime could be obtained from a TeV scale particle decaying through dimension-6 operators suppressed by the grand unification scale $m_{\rm GUT} \sim 2\times 10^{16}$ GeV \cite{Nardi:2008ix, Arvanitaki:2008hq}. Here we point out that such a lifetime can also arise naturally from the goldstino decay.

\begin{figure}[t]
\includegraphics[scale=0.5]{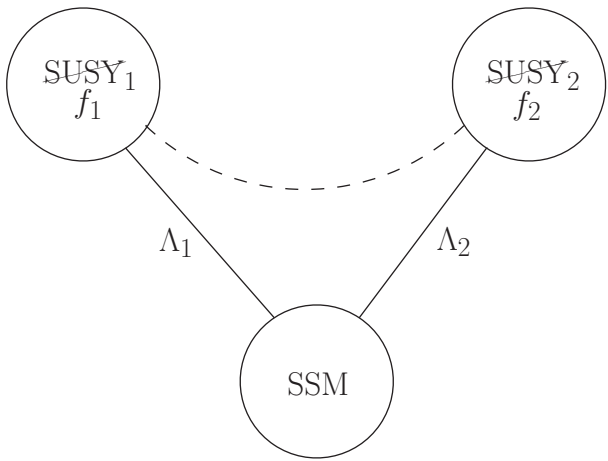} 
\caption{\em \label{fig:fig1} The supersymmetric standard model couples to two sequestered sectors which break SUSY independently through the respective F-term VEVs. The SUSY breaking sectors only directly communicate with each other only through the supergravity effect.}
\end{figure} 

Consider that SUSY is broken by two sequestered sectors, S1 and S2, independently as illustrated in Fig.~\ref{fig:fig1} through the $F$-term VEV's , $f_1$ and $f_2$, respectively. The SSM couples to both sectors and receives soft SUSY breaking mass terms through operators suppressed by energy scales $\Lambda_1$ and $\Lambda_2$. There is one goldstino from each SUSY breaking sector: $\eta_i$, $i=1,2$. One linear combination is eaten and becomes the longitudinal component of the gravitino $\widetilde{G}_L$. The other uneaten combination, $\zeta$, will also acquire a mass due to the supergravity effect, which in the leading order is equal to twice the gravitino mass $m_{\zeta} = 2 m_{3/2}$~\cite{Cheung:2010mc}. Beyond the leading order this relation can be modified \cite{Craig:2010yf}. In our discussion we will assume that the uneaten goldstino is heavier than the gravitino. Assuming $f_1 \gg f_2$, we have $f_{\rm eff} \approx f_1$ and $\widetilde{G}_L$ is mostly $\eta_1$ and $\zeta$ is mostly $\eta_2$. From Eq.~(\ref{eq:Leff}) we see that the four-fermi coupling which governs the goldstino decay into fermions is
\be
\label{eq:goldstino_coupling}
 -\frac1{f_{eff}^2}\left(\frac{\widetilde{m}_1^2 \tan \theta - \widetilde{m}_2^2 \cot \theta}{m_{\widetilde{q}}^2}\right)
 = -\frac1{f_{eff}^2}\frac{f_1}{f_2} \frac{\widetilde{m}_2^2}{m_{\widetilde{q}}^2}\left( \frac{\widetilde{m}_1^2}{\widetilde{m}_2^2} \frac{f_2^2}{f_1^2}-1\right)
 \approx \frac{1}{f_1 f_2} \, \frac{\widetilde{m}_2^2}{m_{\widetilde{q}}^2} \ ,
\ee
 if $\widetilde{m}_1^2$ and $\widetilde{m}_2^2$ are less hierarchical than $f_1^2$ and $f_2^2$: $\widetilde{m}_2^2/\widetilde{m}_1^2 \gg f_2^2/f_1^2$. 
To have the goldstino mass around the TeV scale as the decaying dark matter, $\sqrt{f_1}$ needs to be $\sim 10^{11}$~GeV which is the scale for gravity-mediated SUSY breaking. The goldstino decaying lifetime to a single SM chiral lepton flavor can be estimated from Eq.~(\ref{eq:zgff})
to be,
\be
\tau  \approx 4 \times 10^{26} \ {\rm s} \, \left(\frac{1\,{\rm TeV}}{m_{\zeta}}\right)^9 \left(\frac{\sqrt{f_1}}{10^{11}\  {\rm GeV}}\right)^4 \left(\frac{\sqrt{f_2}}{10^{7}\  {\rm GeV}}\right)^4 \left( \frac{m_{\widetilde{\ell}}^2}{\widetilde{m}^2_{\tilde{\ell} 2}}\right)^2 \left(\frac{0.8}{F_f (x)}\right) ,
\label{eq:lifetime}
\end{equation} 
where $\widetilde{m}^2_{\tilde{\ell} 2}$ is the SUSY-breaking mass contribution to the slepton coming from $f_2$.
We see that the necessary lifetime to explain the PAMELA positron excess can be obtained for $\sqrt{f_2}$ of the order $10^7$ GeV, which is suitable for generating a gauge-mediated contribution to the soft SUSY-breaking masses in SSM.

While PAMELA observed an excess in the positron signals, it did not see any anomalous excess in the anti-proton signals \cite{Adriani:2008zq}. This implies that the decays of the goldstinos should mostly produce leptons, with the hadronic channels not exceeding 10\%~\cite{Nardi:2008ix} if the positron excess is to be explained by the decaying dark matter. From Eq.~(\ref{eq:goldstino_coupling}) we see that the couplings which govern the goldstino decaying to SM fermions are proportional to $ \widetilde{m}_{\tilde{q}2}^2/m_{\widetilde{q}}^2$, the fraction of the soft SUSY-breaking mass of the corresponding superpartner coming from $f_2$. Therefore, the decay into quarks can be suppressed if the squarks have a  smaller fraction of their masses coming from $f_2$ compared with  the sleptons. To satisfy the anti-proton constraint it requires
\be
6\times\left( \frac{\widetilde{m}_{\tilde{q}2}^2}{m_{\widetilde{q}}^2} \right)^2 \ \alt\  0.1\ \left(\frac{\widetilde{m}_{\tilde{\ell}2}^2}{m_{\widetilde{\ell}}^2} \right)^2\qquad \Longrightarrow \qquad \frac{\widetilde{m}_{\tilde{q}2}^2}{m_{\widetilde{q}}^2}\ \alt\  0.13\ \frac{\widetilde{m}_{\tilde{\ell}2}^2}{m_{\widetilde{\ell}}^2}  \ ,
\ee
where the color factor $N_c=3$ is included for the quarks. Such a ratio may result from either a smaller $\widetilde{m}_{\tilde{q}2}^2$ or a larger $m_{\widetilde{q}}^2$. The first possibility could arise if the S2 sector preferentially gives SUSY-breaking masses to the sleptons over the squarks. For example, S2 could couple to the SSM dominantly through the $B-L$ gauge interaction, then it can give 9 times $\widetilde{m}_2^2$ to the sleptons compared to the squarks. In addition, if $f_1$ induces $\widetilde{m}_1^2$ through the usual gravity mediation, the gravity-mediated contributions to the squark masses are generically expected to be much larger then the contributions to the slepton masses due to the running contributions from the gluino mass.\footnote{It is also possible that the contributions to some of the scalar mass-squareds from one of the SUSY-breaking sector are negative. The goldstino couplings to the leptons can be enhanced if the sleptons are light from the cancelation of the mass contributions of the two sectors.}

In addition to suppressing the decays into quarks, the other decay channels which give rise to hadrons such as decays into gauge and Higgs bosons also need to be suppressed. The effective couplings between the goldstini and the gauge or Higgs bosons are presented in the Appendix. The decay widths into these channels  are also controlled by the fractions of the soft masses originated from $f_2$, and hence could be suppressed in similar ways. This can be understood as the uneaten goldstino is mostly composed of $\eta_2$ in the limit $f_1\gg f_2$. In the example of $B-L$ mediation from S2, since the SM gauge bosons and Higgs fields  do not carry $B-L$ charges,  decays to these modes can be even further suppressed than decays to the quarks.\footnote{It is worth pointing out that the coupling to two Higgs fields is particularly dangerous since, after the Higgs field gets a VEV, there is a corresponding two-body decay into  only one Higgs boson plus the gravitino. Typically the phase space of two-body decay is larger than that of the three-body decay by $32\pi^2$, implying that a strong suppression in the goldstino coupling with two Higgs fields is needed.} Another possibility is that  if SUSY breaking in S2 preserves the $R$-symmetry, then  the couplings to the gauge bosons can be naturally suppressed~\cite{Cheung:2010mc}.

An important difference between the goldstino decaying dark matter and many other previously proposed decaying dark matter is that the $R$-parity is exact in this case and there is still a missing particle (gravitino) from the three-body decay. Most previous studies \cite{Chen:2008dh, Nardi:2008ix,Arvanitaki:2008hq} focus on the case of dark matter  decaying 
 to two or four SM particles without any missing particle (other than neutrinos). Since the energy of the decay products is fixed in a two-body decay, if the dark matter decays directly to electrons and positrons, the electron/positron spectrum will exhibit a sharp edge at the half the mass of the dark matter particle even after propagating through the galaxy. The Fermi-LAT measurement of the $e^++e^-$ spectrum does not show any sharp feature below the 1 TeV energy~\cite{Abdo:2009zk}. As a result, the decays dominantly to muons and taus are preferred. On the other hand, in our case the SM particles from the goldstino decay have a smooth spectrum because they come from the three-body decay. The goldstino can decay directly to the electrons/positrons and the energy spectrum can still be consistent with the Fermi-LAT result as we will see in the next section.

In the early universe, the goldstinos can be generated from both thermal productions and decays of superpartners of SM particles. Requiring the correct  relic density for the goldstino dark matter turns out to put strong constraints on the reheating temperature $T_R$. The relic from superpartner decays is expected to be dominated by the sleptons since the goldstino needs to couple most strongly to the sleptons, and the slepton abundance is less Boltzmann suppressed if sleptons are lighter than the other SM superpartners. The decay rate of the slepton to a lepton and a goldstino is \cite{Cheung:2010mc}
\begin{equation}
\label{eq:slepton_decay}
\Gamma_{\tilde{\ell}} = \frac{m_{\tilde{\ell}}}{16\pi} \left( \frac{\widetilde{m}^2_{\tilde{\ell} 1} \tan \theta - \widetilde{m}^2_{\tilde{\ell} 2} \cot \theta }{f_{\rm eff}} \right)^2 \left( 1 -\frac{m_{\zeta}^2}{m^2_{\tilde{\ell}}}\right) \approx \frac{m_{\tilde{\ell}}}{16\pi} \left( \frac{ \widetilde{m}^2_{\tilde{\ell} 2}}{f_{2}} \right)^2 \left( 1 -\frac{m_{\zeta}^2}{m^2_{\tilde{\ell}}}\right) \ .
\end{equation}
Numerically it turns out to be close to the Hubble scale near the typical freeze-out temperature of the slepton,
\begin{equation}
\frac{\Gamma_{\tilde{\ell}}}{H(T)} \approx 0.04 \left(\frac{50\, {\rm GeV}}{T}\right)^2 \left(\frac{10^7\, {\rm GeV}}{\sqrt{f_2}}\right)^4 \left(\frac{m_{\tilde{\ell}}}{1\, {\rm TeV}}\right) \left(\frac{\widetilde{m}_{\tilde{\ell} 2}}{500\, {\rm GeV}}\right)^4 \left( 1 -\frac{m_{\zeta}^2}{m^2_{\tilde{\ell}}}\right).
\end{equation}
This implies that a significant fraction of the sleptons has decayed to goldstinos before the freeze-out. Above the freeze-out temperature the slepton abundance tracks the thermal equilibrium abundance and is exponentially sensitive to the temperature (below the slepton mass). As it is well known that the WIMP miracle means that the amount of a WIMP particle left at its freeze-out temperature is just about right to account for the dark matter if it survives until today. Therefore, the reheating temperature can not be significantly higher than the slepton freeze-out temperature, otherwise there will be too many goldstinos coming from slepton decays, which will over-close the universe. This consideration requires
\begin{equation}
T_R\  \lesssim\  \frac{m_{\tilde{\ell}}}{20} \ .
\end{equation}
Goldstinos can also be produced directly in the thermal bath radiation. From Refs.~\cite{Moroi:1993mb, Cheung:2010mc} we see that if the reheating temperature is higher than the goldstino mass, the goldstino will be over-produced and the parameters in the range of our interest are clearly ruled out. If the reheating temperature is below the goldstino mass, the goldstino production, which is proportional to the square of the radiation number density at the high energy tail, will be suppressed by the Boltzmann factor $\exp(-2 m_{\zeta}/T_R)$. No over-closure of the universe would require
\begin{equation}
T_R \ \lesssim \ \frac{m_{\zeta}}{8} \ .
\end{equation}
If the slepton mass is not much larger than the goldstino mass, this gives a weaker constraint than the constraint from the slepton decays. On the other hand, there is also a lower-bound on the reheating temperature because of the need to produce enough SM superpartners, whose decays result in the right amount of the goldstino dark matter, as in the superWIMP~\cite{Feng:2003xh} scenario. This typically requires the reheating temperature to be higher than the freeze-out temperature $T_F$ of the lightest observable-sector supersymmetric particle (LOSP), which can be a slepton or other superpartners like the neutralino. Combining the constraints together, there is only a small window to achieve the right amount of goldstino dark matter,\footnote{Given that the reheating temperature is below the electroweak scale in the range of parameters we consider, baryogenesis may require non-thermal production of sphaleron configurations~\cite{Krauss:1999ng,GarciaBellido:1999sv}, or come from moduli decays~\cite{Thomas:1995ze,Kitano:2008tk}.}
\be
T_F \left(\sim \frac{m_{\rm LOSP}}{25} \mbox{ for a weakly interacting LOSP} \right) \ \alt \ T_R\ \alt\  {\rm Min}\left\{\frac{m_{\tilde{\ell}}}{20},\ \frac{m_{\zeta}}{8}\right\} \ .
\ee
The fine-tuning required to have the correct relic density seems to be a generic problem for models with two goldstini.

The upper bound of the reheating temperature may be relaxed a little bit if there are more than two independent SUSY-breaking sectors. The PAMELA signals may come from the decay of a heavy goldstino species to a light goldstino species (remembering that their masses can receive corrections to the universal lowest order result). A lifetime similar to Eq.~(\ref{eq:lifetime}) can be obtained with both $\sqrt{f_1}$ and $\sqrt{f_2}$ $\sim 10^9$~GeV, while the overall SUSY-breaking scale $\sqrt{f_{\rm eff}}$ remains at $10^{11}$ GeV due to the presence of additional sectors. Both the direct production of the goldstinos and decays from the SM superpartners in the early universe are then suppressed by a higher scale ($10^9$ GeV). In this case, the upper bound of the reheating temperature may be raised to around $m_\zeta$ itself.

\section{Astrophysical and Collider Phenomenologies}
\label{sect:pheno}

\subsection{Indirect detections}

In this subsection we discuss the implications of dark matter indirect detection of the scenario considered in Section \ref{sect:decayingDM}. As emphasized earlier, this framework differs from the conventional decaying dark matter model in that the dark matter in our case dominantly decays through the three-body process with a missing gravitino. The resulting lepton energy spectrum is softer than that in two-body decays without the missing energy, which allows us to fit both PAMELA positron excess and the lack of sharp edge feature in the Fermi-LAT $e^++e^-$ measurements at the same time, using final states with electrons. In this work we will not be concerned with decays into hadronic final states as well as photons, which are assumed to be suppressed.

We use the Bessel function method of 
Ref.~\cite{Delahaye:2007fr} to calculate the positron flux at the earth, due 
to the decay $\zeta \to \widetilde{G}_L \ell^+ \ell^-$, with the MED model parameters discussed therein, which provide the best fit to the Boron-to-Carbon ratio. For the background fluxes we adopt the ``model 0'' presented by the Fermi-LAT collaboration in Ref.~\cite{Grasso:2009ma}, which are parametrized in Ref.~\cite{Ibarra:2009dr}. As for the dark matter halo model, we use the Moore profile in Ref.~\cite{Diemand:2004wh}. Throughout this section we assume that the dark matter density is $\rho_\odot = 0.3$~GeV/cm$^3$.  We perform  combined fits to both PAMELA and Fermi-LAT data by varying the decay lifetime of the goldstino and the overall normalization of the primary $e^-$ component of the background flux, as described in Ref.~\cite{Ibarra:2009dr}. Moreover, since the $e^+/e^-$ flux at energies below 10 GeV measured at the top of the atmosphere is significantly affected by the solar modulation effect, we only use data points above 10 GeV in the PAMELA measurements to determine the total $\chi^2$.

\begin{figure}
\includegraphics[width=0.7\textwidth]{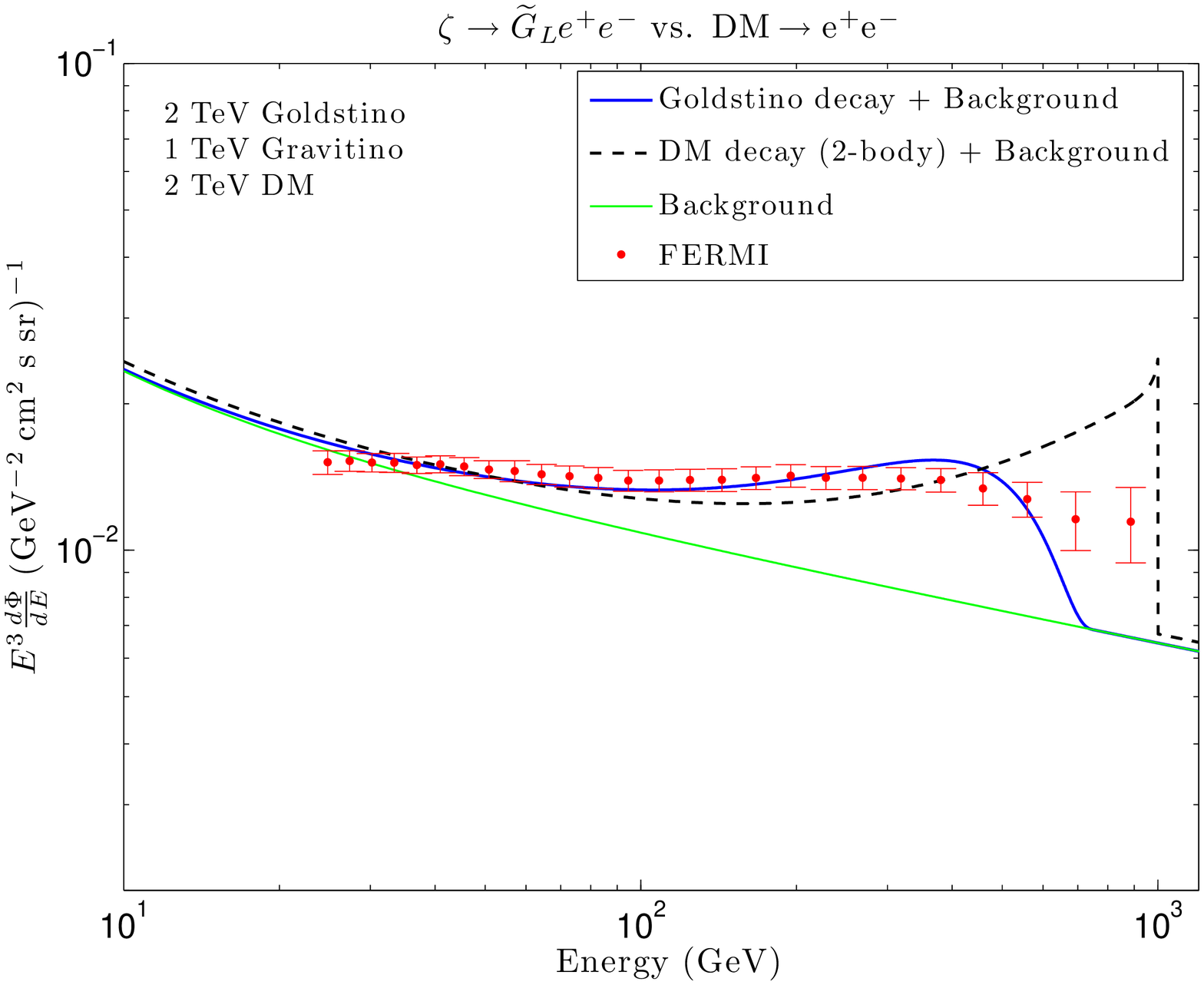}
\includegraphics[width=0.7\textwidth]{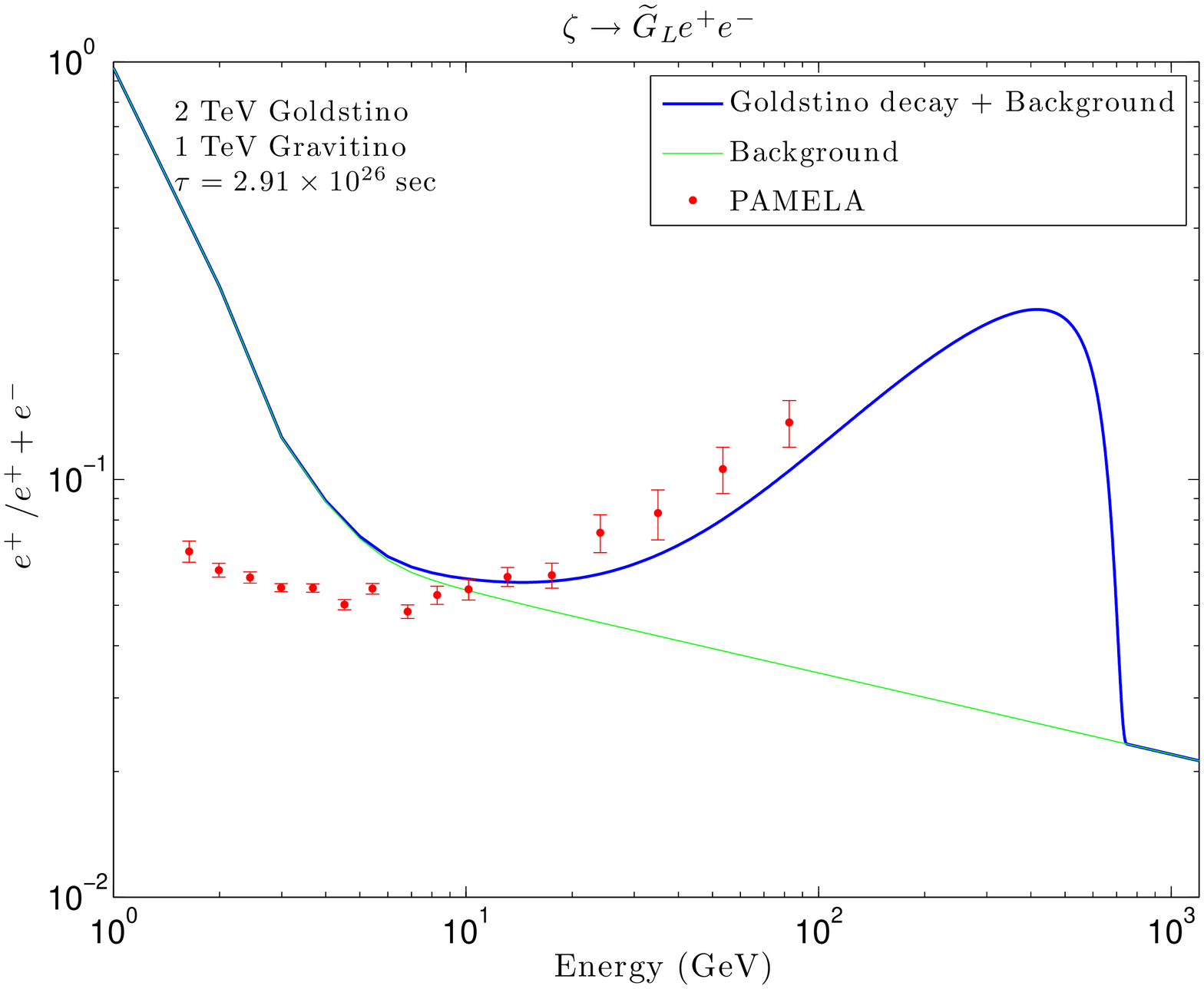}
\caption{\em Fits to the positron fraction and the total $e^++e^-$ flux measured by PAMELA and Fermi-LAT, respectively, using three-body decay of the dark matter (goldstino) into $e^+e^-$ pair together with a missing particle (gravitino). Here we assume the mass relation $m_\zeta=2m_{\widetilde{G}_L}$. The combined $\chi^2$ per degree of freedom is 1.6. We also demonstrate the sharp edge in the two-body decay spectrum in the Fermi fit.}
\label{fig:fitPamelaFermi}
\end{figure}

\begin{figure}
\includegraphics[width=0.7\textwidth]{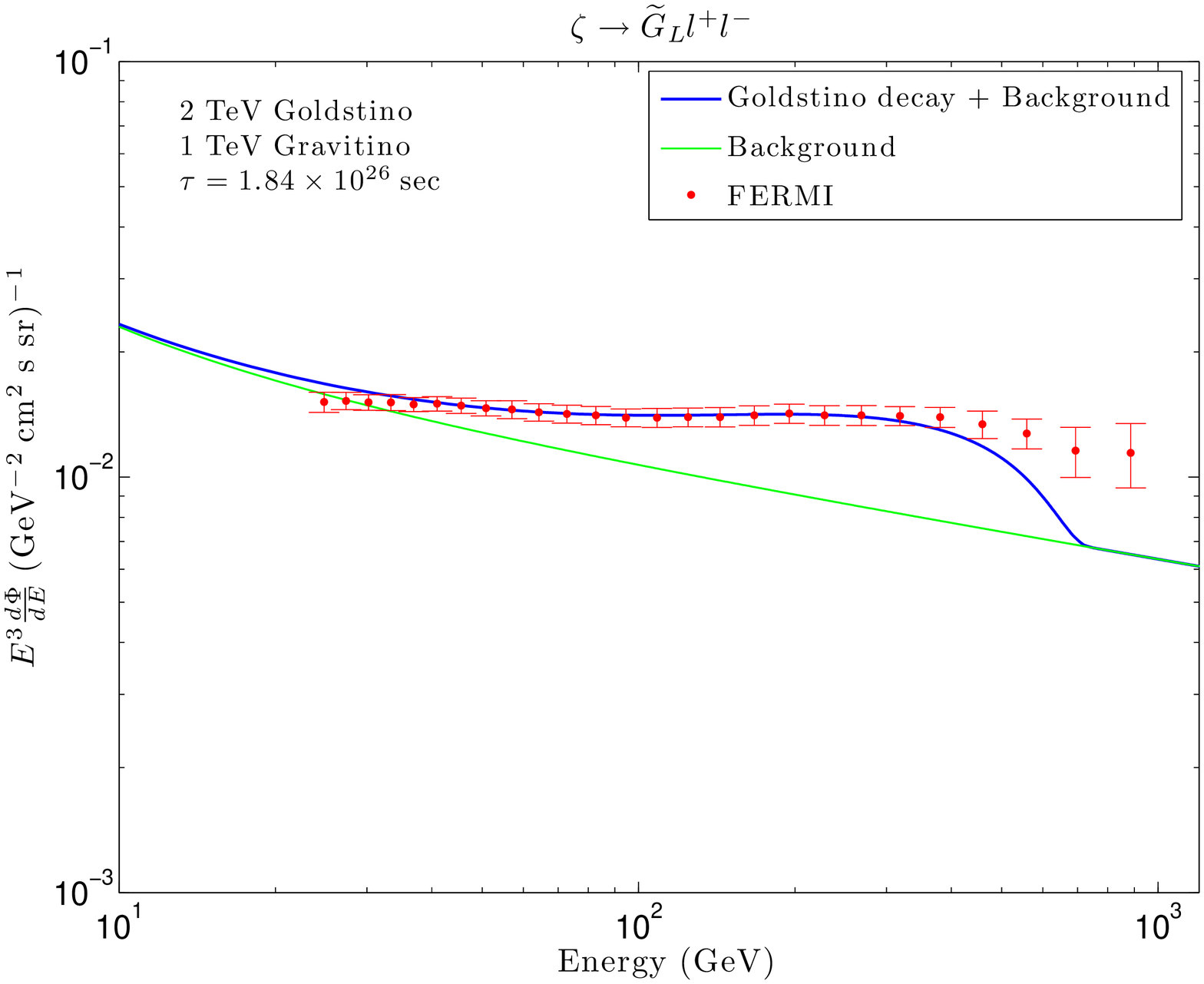}
\includegraphics[width=0.7\textwidth]{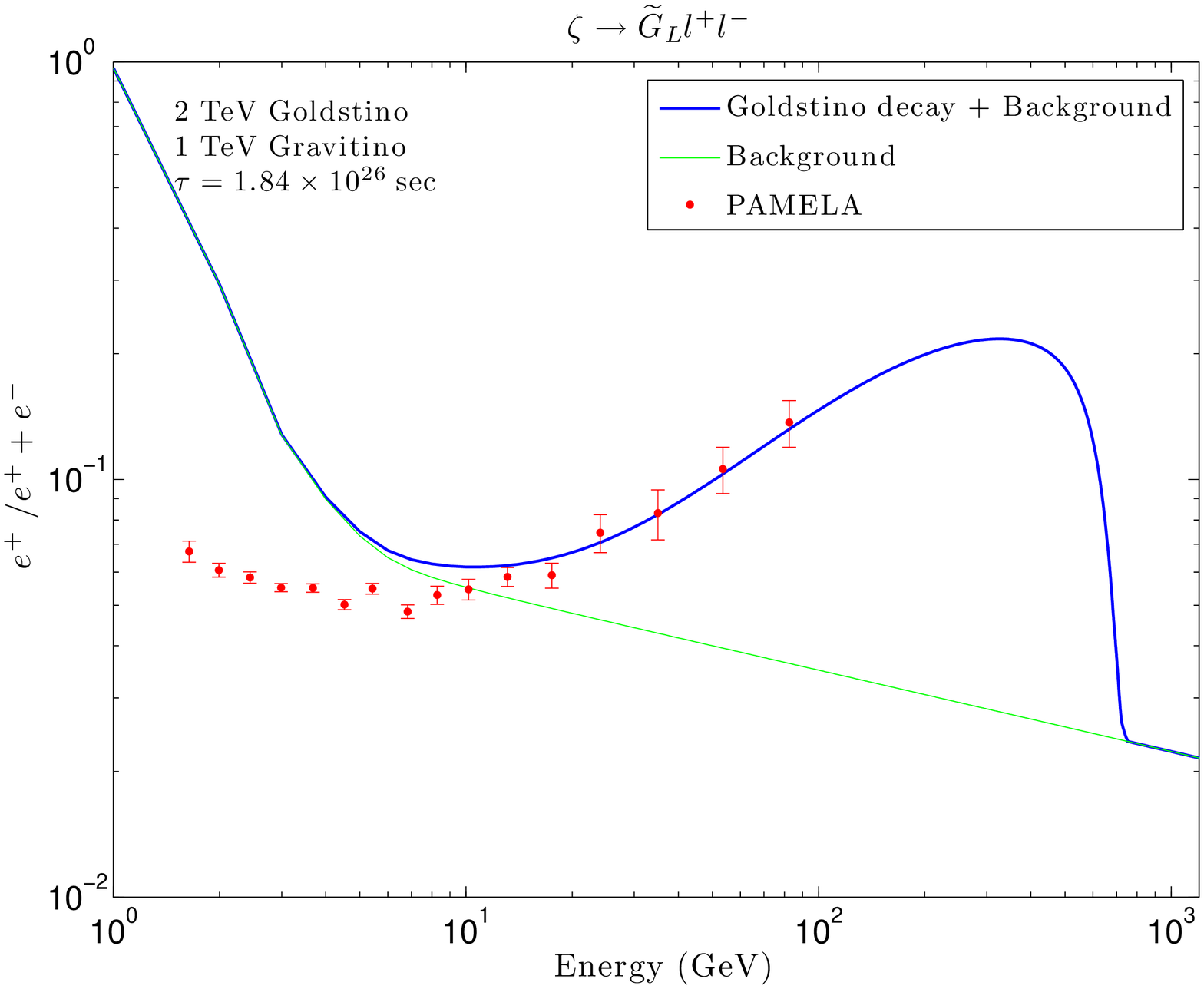}
\caption{\em Fits to the positron fraction and the total $e^++e^-$ flux measured by PAMELA and Fermi-LAT, respectively, using three-body decay of the dark matter (goldstino) into $\ell^+\ell^-$ pair together with a missing particle (gravitino). Here we assume the dark matter coupling to all three lepton flavors is universal. The combined $\chi^2$ per degree of freedom is 1.5. }
\label{fig:fitPamelaFermiuniv}
\end{figure}

In Fig.~\ref{fig:fitPamelaFermi} we show the fit to both the positron fraction of PAMELA and the total $e^++e^-$ flux of Fermi-LAT from a 2 TeV goldstino decay into a gravitino and a pair of $e^+ e^-$ with 100\% branching fraction, assuming the leading order mass relation $m_\zeta=2m_{\widetilde{G}_L}$.  As can be seen, both the rise of positron fraction in the energy regime between 1 and 100 GeV in the PAMELA data and the hardening of spectrum at around 400 GeV in the Fermi-LAT data can be described by the three-body decay into $e^+e^-$ pair plus the missing particle. The smooth feature of energy spectrum resulted from the three-body decay is evident in the figure. In contrast, we also show in the same figure the fit of a two-body decay of dark matter into $e^+e^-$ pair to the Fermi-LAT measurement. The sharp edge at $m_\zeta/2$ is still present even after propagation through the interstellar medium, thus disfavoring this particular decay channel as the explanation for the Fermi-LAT measurement \cite{Meade:2009iu}. In Fig.~\ref{fig:fitPamelaFermiuniv} we demonstrate that reasonable fits to PAMELA and Fermi-LAT can be obtained if the goldstino has universal couplings to all three lepton flavors, which is motivated by the flavor changing constraints on slepton masses.

Some general features of the fit can be understood analytically. The PAMELA data show a rising positron fraction above 10 GeV, while the Fermi-LAT measurements suggest a hardening feature in the region around 400 GeV. For a three-body decay like $\zeta\to \widetilde{G}_L \ell^+\ell^-$, if we neglect the mass of the leptons, the maximum possible energy of $\ell^+$ occurs in the configuration when the lepton and the anti-lepton are collinear, $q^2\equiv (p_{\ell^+}+p_{\ell^-})^2=0$, and $E_{\ell^-}\to 0$. In this case, conservation of momentum gives $p_{\widetilde{G}_L} = p_\zeta-q$ and hence
\be
\label{eq:endpoint}
E_{\ell^+}^{\rm (max)} = \frac{m_\zeta^2-m_{\widetilde{G}_L}^2}{2m_\zeta} < \frac{m_\zeta}2  
\ee
in the $\zeta$ rest frame.
The position of  the Fermi-LAT plateau set a lower bound on the end point of the lepton energy, if we were to explain it with a signal component in the observed flux, 
\be
E_{\ell^+}^{\rm (max)} \agt {\cal O}( 400 \ {\rm GeV}) \quad \Longrightarrow \quad m_\zeta \agt  {\cal O}( 800 \ {\rm GeV}) \ ,
\ee
from which we conclude that it will be difficult to fit Fermi-LAT measurement with a dark matter mass less than  1 TeV, an observation that has been reached previously \cite{Meade:2009iu}. 

\begin{figure}[t]
\includegraphics[width=0.7\textwidth]{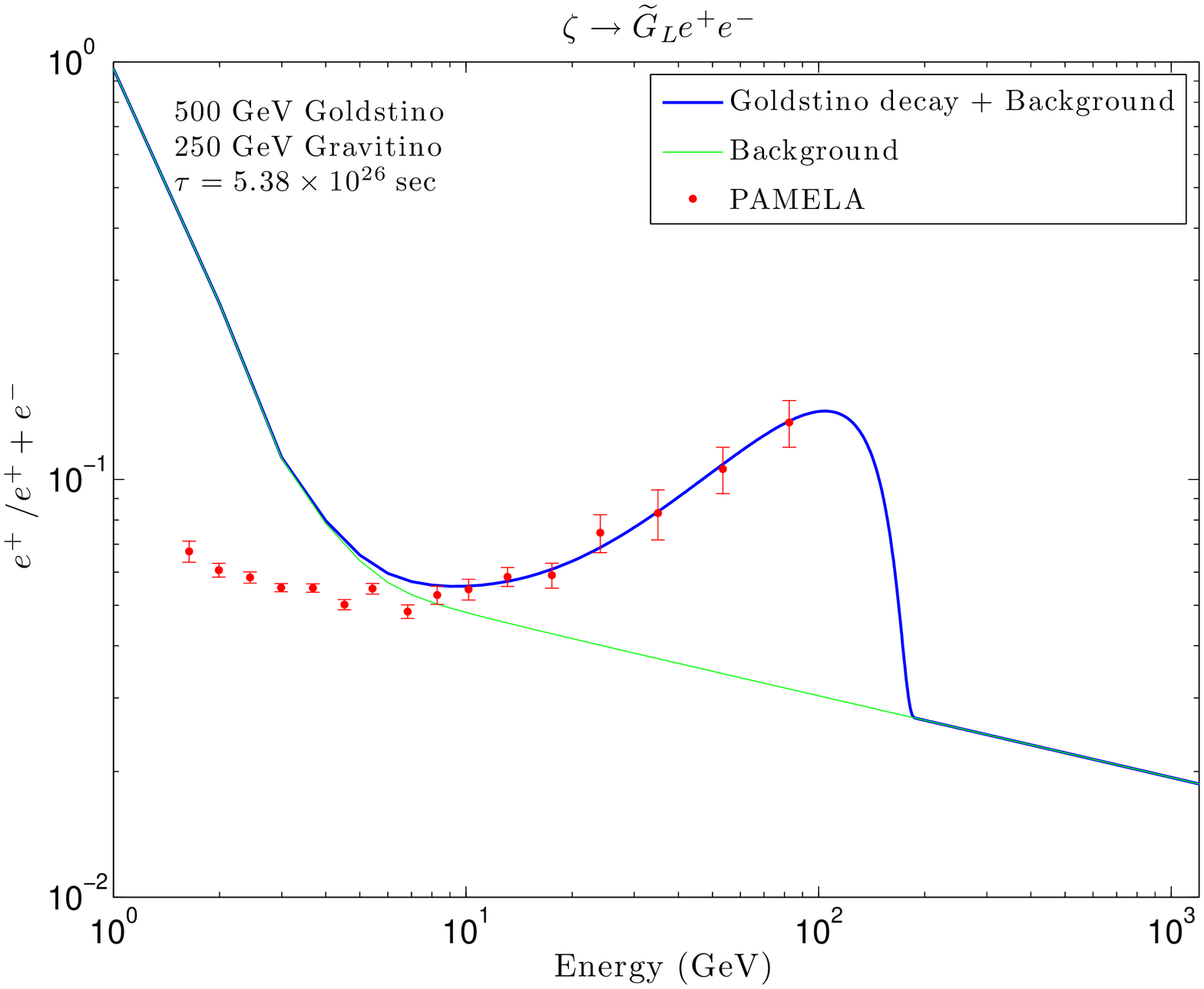}
\includegraphics[width=0.7\textwidth]{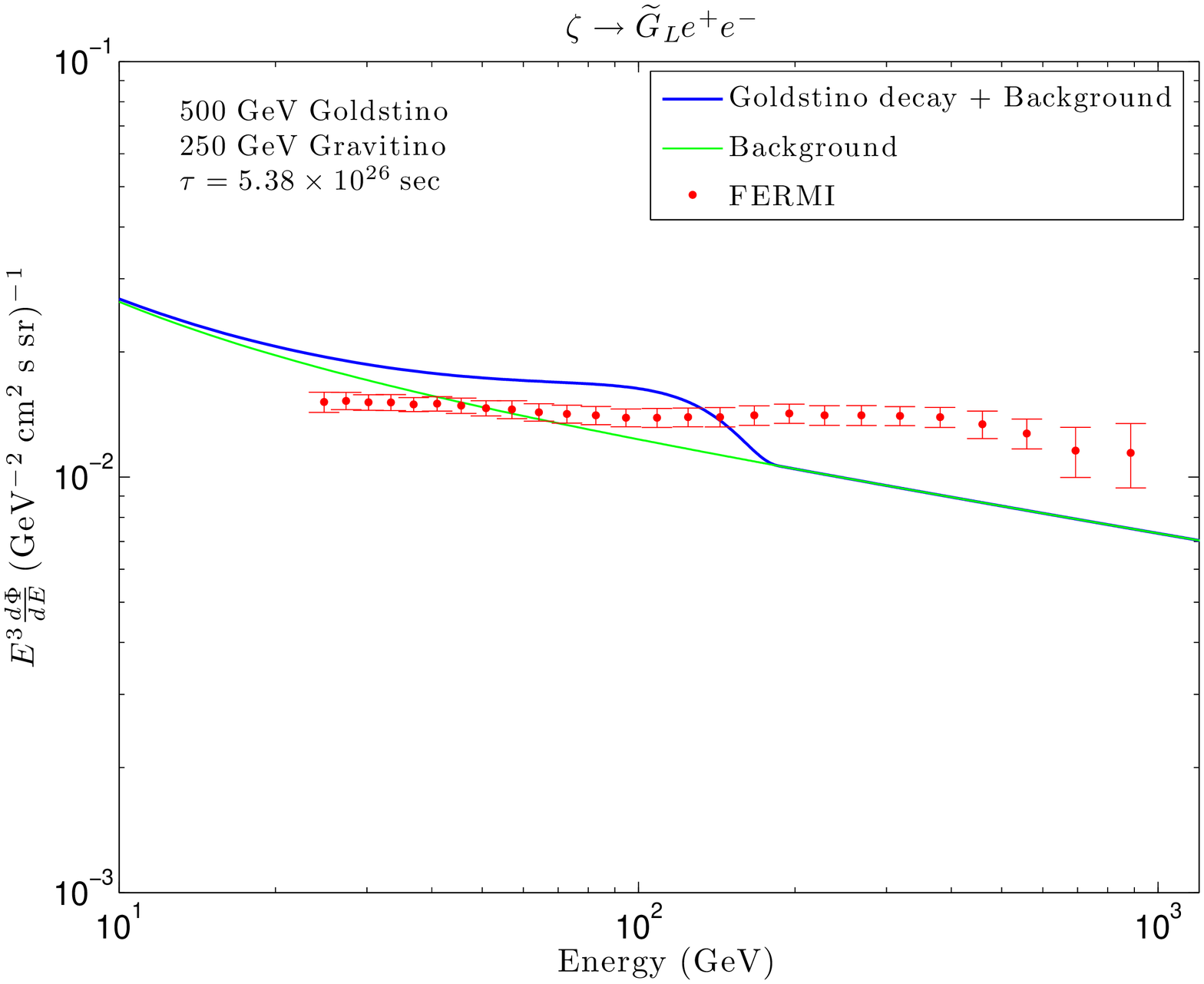}
\caption{\em Fit to the positron fraction measured by PAMELA using a 500 GeV goldstino, and the resulting signal flux when added on top of the Fermi-LAT background. In this case, the hardening of Fermi-LAT spectrum needs to be explained by some other sources.}
\label{fig:fitPamelaFermi500TeV}
\end{figure}

The conclusion about the heaviness of the dark matter seems quite robust against different choices of background fluxes. For example, the Fermi-LAT collaboration provided two other backgrounds, in addition to the ``model 0'' background adopted in this work, which in fact give good fits to their data even in the absence of any exotic sources of $e^++e^-$ flux. These backgrounds, model 1 and model 2 in Ref.~\cite{Grasso:2009ma}, do not provide good fits to pre-Fermi data measured by other experiments. We studied the possibility of using these other background to perform the fit to PAMELA and Fermi-LAT. The resulting fits are much worse in the case of PAMELA data, since the signal flux that can be accommodated by the Fermi-LAT data is too small to explain the rise in the positron fraction.

The PAMELA anomaly alone can be fitted with a much lighter goldstino, if there are other sources which can account for the hardening of the Fermi-LAT spectrum around 400 GeV. In Fig.~\ref{fig:fitPamelaFermi500TeV} it is shown that a 500 GeV decaying goldstino can fit the PAMELA positron fraction very well. However, the deficit of the Fermi-LAT spectrum above 100 GeV needs to be explained by some other sources, such as those discussed in Refs.~\cite{Kane:2009if,Hooper:2009cs}.

\subsection{Collider phenomenology}

The collider phenomenology of the scenario, that the gravitino and goldstino are the LSP and NLSP,  has been discussed in Ref.~\cite{Cheung:2010mc,Cheung:2010qf}. Here we only give a brief summary, focusing on the parameter region in which we are interested. The SM superpartners produced at the collider will cascade-decay down to the LOSP. The LOSP will travel some distance before decaying to the goldstino. Therefore, the collider signals depend on which SM superpartner is the LOSP. Because the goldstino couples most strongly to the leptons, the most natural candidate for the LOSP is the slepton in this case. The decay length of a slepton LOSP is estimated to be
\begin{equation}
c\, \tau \approx 1.6\, {\rm m}\ \left( \frac{\sqrt{f_2}}{10^7\, {\rm GeV}} \right)^4 \left(\frac{1\, {\rm TeV}}{m_{\tilde{\ell}}}\right) \left(\frac{500\, {\rm GeV}}{m_{\tilde{\ell}_2}}\right)^4 \left(1- \frac{ m_\zeta^2}{m^2_{\tilde{\ell}}}\right)^{-1}.
\label{eq:losplength}
\end{equation}
If the long-lived LOSP is a charged slepton, the collider signature is very distinct. Measuring the charged track can determine the mass of the long-lived particle. In addition, a large fraction of them will decay inside the detector, leaving a displaced kink in the tracking detector, which allows a measurement of the LOSP lifetime. On the other hand, if the LOSP is a sneutrino, its decay is invisible and there is no distinct feature other than the usual missing energy signals for the SUSY events.

The LOSP may be other superpartners if their couplings to the goldstino is much suppressed relative to the slepton couplings to the goldstino. The lifetime of the LOSP in this case needs to be longer than that given in Eq.~(\ref{eq:losplength}). For a colored LOSP (gluino or squark), it will hadronize and form $R$-hadrons; see Ref.~\cite{Fairbairn:2006gg} for a review on the experimental searches. A fraction of the $R$-hadrons could be stopped in the detector and decay later, resulting in distinctive signatures \cite{Arvanitaki:2005nq}. If the LOSP is a neutralino, it will escape the detector most of the time, giving rise to the standard missing energy signals for SUSY. However, a small fraction of the neutralinos will decay inside the detector, producing $\gamma$, $Z$, or $h$. These decays can be discovered if the lifetime is shorter than $10^{-3}$--$10^{-5}$ second \cite{Ishiwata:2008tp,Chang:2009sv}. If the direct LOSP coupling to the goldstino is highly suppressed, the LOSP decay to the goldstino may be dominated by 3-body process through the off-shell sleptons. If this indeed happens, it provides a nice check that the goldstino couples mostly to the sleptons.

\section{Conclusions}

\label{sect:conclusions}

In this work we proposed a new scenario for supersymmetric decaying dark matter in theories with goldstini, where the uneaten goldstino dominantly decays into gravitino, which shows up as missing energy, and two SM particles. In this scenario it is not necessary to introduce $R$-parity violations, since the goldstino decays through dimension-8 operators and naturally has a long lifetime suitable to explain the positron excess observed by the PAMELA collaboration. We derive low-energy effective interactions of the goldstini and show that the couplings can be non-universal, while the gravitino coupling remains universal as expected. The non-universality of the goldstini coupling is crucial for the dark matter to be leptophilic, so as to avoid the lack of excess in the anti-proton spectrum measured by PAMELA. To obtain the correct goldstino relic density for the dark matter, however, seems to require some fine tuning of the reheating temperature in the early universe. 

A distinct feature of this scenario is the three-body decay of the dark matter, which results in softer energy spectra for the electrons and positrons, as opposed to a sharp edge in the case of the more conventional two-body decay. Consequently, it is possible to fit both the positron excess in the PAMELA data and the hardening feature in the $e^++e^-$ flux measured by the Fermi-LAT. We find decays into $e^++e^-$ with 100\% branching fraction, which is disfavored if the dark matter decays  into two or four SM particles, could still provide reasonably good fits to PAMELA and Fermi-LAT. In addition, universal coupling of the dark matter with all three lepton flavors, which may be favored from other considerations, could also fit the data well.

In this work we have assumed the hadronic decay modes of the dark matter, as well as prompt decays into photons, are suppressed in order to satisfy constraints from anti-proton and gamma ray measurements. However, it is worth pointing out that most studies on these constraints are based on the assumption that the dark matter decays into two-body final states, while the decay proceeds through three-body channel with a missing particle in our scenario. It would be interesting to re-evaluate these constraints in a more model-independent fashion for the case of three-body decays with missing particles.

\begin{acknowledgments}
We would like to thank Spencer Chang and Carlos Wagner for useful discussion. We are also grateful to Gabe Shaughnessy and Shashank Shalgar for assistance in generating the figures. H.-C.~C. thanks the hospitality of Northwestern University, where this work was initiated, and Fermilab, where part of this work was performed. This work is supported in part by the U.S.~Department of Energy under contracts DE-AC02-06CH11357,  DE-FG02-91ER40684,  DE-FG02-94ER40840, and  DE-FG02-91ER40674.
\end{acknowledgments}

\appendix*

\section{Effective Interactions of Goldstini}
Here we present goldstini couplings with the gauge and Higgs bosons without detailed derivations, which are beyond the scope of the current work and will be presented elsewhere \cite{mssmlhc}. The goal is to demonstrate that, in the limit $f_1\gg f_2$, these couplings are proportional to the fraction of soft masses coming from $f_2$.  We use 2-component spinors throughout the Appendix.


Effective interactions of goldstini with two $U(1)$ gauge bosons are derived by using the following UV interactions:
\be
\int d^2\theta \left(\frac1{4} + \sum_{i=1,2} \frac{1}{2\Lambda_i}X_i\right) W_\alpha W^\alpha + {\rm h.\, c.} \ ,
\ee
where the field strength superfield $W_\alpha$ is
\bea
W_\alpha &=& -i \lambda_\alpha+L_\alpha^\beta \theta_\beta
     +\sigma^\mu_{\alpha\dot{\alpha}} \partial_\mu \ov{\lambda}^{\dot{\alpha}}\, \theta^2 \ , \\
L_\alpha^\beta &=& \delta^\beta_\alpha D - {i} F^\beta_\alpha    \ .
\eea
In the above $F(\ov{F})\equiv F_{\mu\nu} \sigma^{\mu\nu} (F_{\mu\nu} \ov{\sigma}^{\mu\nu})$.  Similar to the case of four-fermi interactions,  there are two contributions to the two-goldstino and two-photon interactions after integrating out the gaugino,  arising from the zero momentum lagrangian and the gaugino kinetic term, respectively,
\bea
{\cal L}_{2\gamma}^{(0)} &=&-\frac{i}{\sqrt{2}} \frac{f_{eff}}{f_2\Lambda_1+f_1\Lambda_2} \ov{\eta} \ov{F} \, F \eta
 = \frac1{m_\lambda}\left(\frac{\widetilde{m}_1}{\Lambda_1}+\frac{\widetilde{m}_2}{\Lambda_2}\right) \ov{\eta} \ov{F} \, F \eta
  \ ,  \\
{\cal L}_{2\gamma}^{(1)} &=&  \frac{i}{2 f_{eff}^2}\left[\, \ov{\widetilde{G}}_{L}\, \ov{F}\, \sigma \cdot \partial  \left(F\, \widetilde{G}_{L} \right) +\left( \frac{\widetilde{m}_1\tan\theta -\widetilde{m}_2 \cot\theta }{m_{{\lambda}} }\right) ^2\ov{\eta}\, \ov{F}\, \sigma \cdot \partial  \left(F\, \eta \right) \right. \nonumber  \\
&&\left.- 2\left( \frac{\widetilde{m}_1\tan\theta -\widetilde{m}_2 \cot\theta }{m_{{\lambda}} }\right)\ov{\widetilde{G}}_{L}\, \ov{F}\,\sigma \cdot \partial  \left(F\, \eta \right) \right] \ ,
\eea
where $\widetilde{m}_i=f_i/\Lambda_i$ and $m_\lambda=\widetilde{m}_1+\widetilde{m}_2$.  Again we see that the gravitino effective interaction is universal while those involving the goldstino are not. There is also a non-derivative coupling for the goldstino.

There is also a three-point coupling contributing to two-body decays of a goldstino into the gravitino and one massive gauge boson, which is nonetheless suppressed by the $D$-term \cite{Komargodski:2009rz, Luty:1998np}. It is very small and will not be considered here.

For couplings with the Higgs bosons we consider the following K\"ahler potential and superpotential:
\bea
K &=&  \sum_{i=1}^2 \left(X_i^\dagger X_i - \frac{c_i}{\Lambda^2}
\left(X_i^\dagger X_i \right)^2 \right) +
\sum_{i=1}^2 \sum_{\alpha=u}^d \left(1 - \frac{g_{i\alpha}}{\Lambda_i^2} X_i^\dagger X_i \right)
H_\alpha^\dagger H_\alpha \\
W &=& \sum_{i=1}^2- f_i X_i + \mu \left(1+ \frac{d_i}{\Lambda_i} X_i \right) H_u H_d \ ,
\eea
We will work in the limit $\Lambda_i \gg \Lambda$ so that  $\widetilde{x}_i = {\eta_i^2}/(2f_i)$ as before. In addition, we define the soft masses $B_i = {d_i f_i}/{\Lambda_i}$ and $m_{i\alpha}^2 ={g_{i\alpha}f_i^2}/{\Lambda_i^2}$ such that
\be
\mu \sim B_i  \sim m_{soft} \quad {\rm and } \quad m_{i\alpha}^2  \sim m_{soft}^2 \ ,
\ee
and only keep contributions up to ${\cal O}(m_{soft}^2/f_i)$. Then the effective interactions relevant for goldstino decay into two Higgs bosons are
\bea
{\cal L}_{2h}^{(0)} &=&-\frac{1}{ \mu f_{\rm eff}^2} 
\widetilde{G}_L \zeta \left[\left(m_{H_u}^2 + |\mu|^2 \right) 
\phi_u^\dagger- B\mu \phi_d\right] \left[\delta m_d^2 \phi_d^\dagger-
\delta B\mu \phi_u \right] \nonumber \\
&&  + u \leftrightarrow d  + \rm{h.c.} \, , \\
{\cal L}_{2h}^{(1)} &=& \frac{1}{\mu^2 f_{\rm eff}^2} \left[\partial_\mu \left\{\left(\left(m_{H_u}^2 
+ |\mu|^2 \right) \phi_u- B\mu \phi_d^\dagger\right) \ov{\widetilde{G}}_L 
\right\}\, i\ov{\sigma}^\mu\, \left(\delta m_{u}^2 \phi_u^\dagger - \delta B\mu \phi_d \right) \zeta 
\right. \nonumber \\
& & \left. + \partial_\mu \left\{\left(\delta m_{u}^2 \phi_u - \delta
B\mu \phi_d^\dagger \right) \ov{\zeta}
\right\} \, i\ov{\sigma}^\mu\,  \left(\left(m_{H_u}^2 
+ |\mu|^2 \right) \phi_u^\dagger- B\mu \phi_d\right) \ov{\widetilde{G}}_L\right]  \nonumber \\
&&  + u \leftrightarrow d  + \rm{h.c.} \ ,
\eea
where 
\bea
m_{H_\alpha}^2 &=& \sum_i m_{i\alpha}^2 \ , \quad \delta m_\alpha^2 =m_{1 \alpha}^2 \tan \theta - m_{2\alpha}^2 \cot \theta \ , \quad \alpha= u, d \ , \\
B &=& \sum_i B_i\  \ \ , \  \ \quad \delta B = B_1 \tan \theta - B_2 \cot \theta \ . 
\eea
It can be seen by the equation of motion for the Higgs fields,
\bea
&& \left(m_{H_u}^2 + |\mu|^2 \right) 
\phi_u^\dagger- B\mu \phi_d = \Box \phi_u^\dagger \ , \\
&&  \left(m_{H_d}^2 + |\mu|^2 \right) 
\phi_d^\dagger- B\mu \phi_u = \Box \phi_d^\dagger \ ,
\eea
that the above interactions are derivatively coupled, as should be for the gravitino coupling.


\end{document}